\documentclass[endfloats]{jpsj2}
\usepackage[abs]{overpic}
\usepackage{booktabs}
\usepackage{cite}
\usepackage{graphicx}

\title{Electron Transfer from Hydrogen Molecule to Au(111) During \\
Dissociative Adsorption: A First-Principles Study}

\author{Shigeyuki \textsc{Takagi},
Jun-ichi \textsc{Hoshino},
Hidekazu \textsc{Tomono}, and
Kazuo \textsc{Tsumuraya}
\thanks{E-mail address: abinitio@isc.meiji.ac.jp}}

\inst{Department of Mechanical Engineering Informatics, School of Science and Technology, \\
Meiji University, Kawasaki 214-8571, Japan}

\abst{
We investigate the electron transfer from a dissociatively adsorbed
H$_2$ molecule to a Au(111) surface using the first-principles methods.
A fractional electron transfers from a molecule to a substrate,
and potential energy increases during the process.
The initial energy increase coincides with that of the isolated, separated, and
positively charged H$_2$ molecule calculated by the real-space density
functional method.
The barrier formation is due to the destabilization of the molecule
induced by the electron transfer.
The electronegativity difference between the adsorbate and the substrate 
determines the direction of the electron transfer. }

\kword{electronegativity difference, dissociation, direction of electron
transfer, adsorption, hydrogen molecule, surface, electron transfer, energy
barrier }

\begin{document}
\maketitle
\newpage
\section{Introduction}
Understanding and controlling the reactivity of adsorbed molecules are one of
the central issues in the field of surface science. 
The term "reactivity" refers to surfaces's ability to break bonds of
molecule approaching surfaces and to adsorb fragments that are forming
a new bond with atoms on the surfaces.
Bond breaking is a rate-limiting step in catalytic reactions.
If there is no barrier during the adsorption, then a molecule has high
reactivity when fragments have large chemisorption energies;
if there is a barrier, then a molecule has low reactivity.
Among the adsorptions of molecules on metal surfaces, the adsorption
of H$_2$ molecules on metals is a typical example of catalytic
reactions, since they have been used in fuel cells.
The adsorption barriers have been confirmed theoretically to exist for the
H$_2$ adsorption on simple metal surfaces such as Al\cite{hamm3,joha}
and Mg\cite{nobu,nors1,joha} and on the noble metals such as
Cu,\cite{hamm2,hamm3,hamm4,harr,hamm5,baer,dino} Ag,\cite{eich} and 
Au\cite{hamm2,stro,hamm5} surfaces with filled $d$-orbitals.
No barrier has been confirmed on the Ni,\cite{hamm2,nobu,kres,harr,hamm5,dino}
Pd,\cite{wei,nobu,dong} Pt,\cite{hamm2,hamm3,olse,hamm5} and W\cite{holl,whit}
surfaces that have unfilled $d$-bands.

The origin of the formation of these barriers has been first given by
N{\o}rskov's group.\cite{nors1,joha,nors2}
They explained the barriers to be due to kinetic energy repulsion,
that had proposed by Zaremba and Kohn,\cite{zare} to explain the interaction
between closed shell adsorbates, such as isoelectronic He atoms, and
metal surfaces. 
Harris and Anderson,\cite{harr} who have first used Pauli repulsion for the
barrier formation, showed that a barrier appears in a Cu$_2$H$_2$ cluster and
no barrier appears in a Ni$_2$H$_2$ cluster.
They explained that since the atoms with $d$-holes such as the Ni atom accept
an $s$-electron from the H atom, no barrier appears by canceling the Pauli repulsion.
A copper atom has the filled $d$-orbitals, so the Pauli repulsion produces the barrier.
Feibelman $et$ $al$.\cite{feib} and Hammer $et$ $al$.\cite{hamm3,hamm2} have
also shown the barriers of the H$_2$ molecule on the Al and Mg metal
surfaces and on the Cu, Ag, and Au metal surfaces to be attributed to the Pauli
repulsion and no barrier for the Ni, Pd, and Pt surfaces to the cancellation.
Hammer $et$ $al$. have also explained the barriers for the Al and Mg metal
surfaces to be due to the "weak chemisorption" case reported by Newns
and Anderson.\cite{ande,newn}

In 1995, N{\o}rskov $et$ $al$. have given an explanation for the barrier
formation using projected density of states.\cite{hamm3,hamm2}
On the one hand, when both the bonding and the anti-bonding states are
situated below the Fermi level, then the barrier appears owing to the Pauli
repulsion.
This is the case for the Cu and Au metal surfaces. 
On the other hand, when the Fermi level is located between the anti-bonding
and the bonding states, the hybridization is attractive, counteracting
the Pauli repulsion and deleting the energy barrier.
These are the cases for the Ni and Pt surfaces that have $d$-band holes.\cite{hamm2}

Since the extensive discussion of the explanations for the barrier formation
from 1981 to 1995, all first-principles studies have explained the barrier
formation using both the Pauli repulsion and the $d$-band hole.
\cite{zare,nors1,joha,nors2,feib,hamm3,hamm2} 

There have been only two studies on the difference electron density of the
dissociated H$_2$ molecules on the metal surfaces.\cite{bird,huda}
Bird $et$ $al$. have shown an increase of the electron density of
the dissociated H$_2$ molecule on Mg(0001) surface through the
difference electron density map.\cite{bird}
Huda and Ray have shown electron transfer from the Pu(111) surface to the
H$_2$ molecule.\cite{huda}

The aim of the present study is to investigate how the electron population
varies during the process and to elucidate the relation between the electron
transfer and the barrier formation.
There has been no earlier study, to the authors's best knowledge, 
that has investigated such a relation.

We evaluate electron density using the first-principles calculations
for the dissociation of the H$_2$ molecule on a Au(111) surface.
There have been earlier studies of the dissociative adsorption of
the H$_2$ molecule on the periodic Au surfaces.\cite{hamm2,corma2007}
This system has been found to have a large dissociation barrier.\cite{hamm2}

\section{Computational Details}
We perform first-principles density functional calculations using the
generalized gradient approximation of Perdew, Burke and Ernzerhof
(GGA-PBE).\cite{pbe}
The interactions between the ion cores and electrons are described using
the ultrasoft pseudopotential\cite{uspp} for Au atoms and the
norm-conserving pseudopotential\cite{ncpp} for H atoms.
We use the valence $5d^{10}6s^{1}6p^{0}6d^{0}$ configuration with partial
core correction for Au pseudoatoms and the valence $1s^{1}2p^{0}3d^{0}$
configuration for H atoms.
A partial core correction is necessary for partitioning the space with
the minimum electron densities around each atom (see below).
The cutoff radius of the partial core for Au atoms is 0.582\AA.
Brillouin-zone integration is performed using a Monkhorst-Pack 
grid\cite{monk} of 4$\times$3$\times$1 {\mbox{\boldmath $k$}}-points.
We use a planewave-based pseudopotential formalism with a periodic boundary
condition to calculate the electronic structures of the dissociation
of the H$_2$ molecule.

We use a slab structure to model the surface of the Au bulk system.
The surface consists of a $\sqrt{3}\times2$ structure with four Au atoms
and the slab contains four Au layers.
Figure \ref{fig1}(a) shows the geometry of the H$_2$/Au(111) system that we
investigated and the numbers of Au atoms on the first layer. 
The integrated electron charge variation, as we will show later, within the
second layer has been negligible during the dissociative adsorption.
Thus this number of layers is sufficient for the present purpose of calculation.
The slab structure is sandwiched by two vacuum layers whose
thickness is 6.59 \AA, which corresponds to the vacuum region with 13.18 \AA.
We select the origin of the reaction path at Z=4\AA ~ above the first
layer, at which the binding energy between the H$_2$ molecule and the
Au surface has been only 0.0247 eV which is sufficiently low compared
with the maximum barrier height of 1.55 eV, as will be shown in Fig. \ref{fig3};
the present size of the vacuum region is sufficient for the present
analyses.
The slab structure has been relaxed;
the distance between the first and second layers is 2.615 \AA,
and the distance between the second and the third layers is 2.603 \AA.

We determine the cutoff energy of the planewave expansion of the
wavefunction for the Fourier expansions of both the
psudowavefunctions and the psudopotentials to converge.
The cutoff energy for the electron density is chosen to be nine times the
value of the wavefunction to incorporate the partial core charge.
The number of basis functions decreases with increasing cell size at a
fixed energy cutoff, which increases total energy with increasing cell
size leading to an incorrect equation of states.
Thus we calculated the equation of states for a fixed number of bases.
The cutoff energy is chosen to be 27.79 Ry at 3.98 \AA ~of cell size.
The maximum barrier height has increased only by 0.039 eV at an increased
cutoff energy of 40 Ry. Thus, the selected cutoff energy is sufficient for the
present analyses. 
The H$_2$ molecule is located above the top site of the Au0 atom with
its molecular axis kept parallel to the surface and is oriented
towards the bridge sites, as shown in Fig. \ref{fig1}(b).
To calculate the potential energy surface of the dissociated
H$_2$ molecule on the slab structure, we constrain the atom positions
of the whole system.

To evaluate the electron transfer between the H atoms and the surface
atoms, we use Bader analysis,\cite{bader} in which the space is partitioned
along minimum-charge-density surfaces, which is called zero-flux planes
around the atoms. By integrating the charges inside each region, we evaluate
the electrons that belong to each atom and then evaluate the electron transfer
through the difference from those of reference systems. 
We have confirmed the existence of the zero-flux planes in the present
investigation except for the plane between the H atoms in an undissociated 
equilibrium H$_2$ molecule, since they form covalent bonding. 
Thus, we integrate the electron densities around the two H atoms to show
the electron transfer from the molecule to the surface.
We compare the stabilities of the dissociatively adsorbed and charged H$_2$
molecule in the whole system with those of the isolated, separated, and
charged H$_2$ molecules in real-space.
We calculate the energies of the molecule using a real-space
density functional method with the same electron-correlation functionals
as the periodic system.

We use the PHASE code\cite{phase} for the plavewave calculations and the
Amsterdam Density Functional (ADF) code\cite{adf} for the real-space calculations
using Slater-type orbitals with triple-zeta basis sets and doubly polarized basis
functions, which are the same conditions as those used in our previous
investigation for the hydrated silicon cluster.\cite{take}

\section{Results}
Figure \ref{fig2} shows the potential energy surface for the dissociatively
adsorbed H$_2$ molecule.
The energies are shown as a function of the bond length $d$ and the height
$Z$ of the molecule above the surface.
A reaction path is defined in the potential energy surface.
The barrier top is located at a height of 1.5 \AA ~above the top layer of
the slab with a H-H distance of 1.5 \AA.
The top shown with an open triangle mark appears after the molecule has
dissociated, since the equilibrium H$_2$ distance is 0.74 \AA.
The open square mark in the figure will be discussed in 
Figs. \ref{fig5}-\ref{fig7}. 

In order to show the potential energy variation $\varDelta E$ along the
path $s$, we use the equation
\begin{equation}
\label{eq:eq1}
\varDelta E = E[\mathrm{H}_2,\mathrm{Au}(111)] - E[\mathrm{H}_2] - E[\mathrm{Au}(111)],
\end{equation}
where the first term is the formation energy of the whole system and the other
terms are the energies of the H$_2$ molecule and the Au(111) slab
structure in the supercell with the same sizes as the first term.
The first term is expressed by
\begin{equation}
  E[\mathrm{H}_2,\mathrm{Au}(111)]
               = E_T[\mathrm{H}_2,\mathrm{Au}(111)] - 2\mu_{\mathrm{H}} - 16\mu_{\mathrm{Au}},
\end{equation}
where the first term is the total energy of the whole system and the other
ones are the chemical potentials of the H and Au atoms. Inserting the same type
equations for the other terms into the eq. (\ref{eq:eq1}), we obtain the following
equation for the potential energy variation;
\begin{equation}
  \varDelta E = E_T[\mathrm{H}_2,\mathrm{Au}(111)] - E_T[\mathrm{H}_2] - E_T[\mathrm{Au}(111)],
\end{equation}
This corresponds to the energy zero to be selected for the H$_2$
molecule to be located at an infinitely separated position above the surface.
The energy $\varDelta E$ along the path $s$ is shown in Fig. \ref{fig3}.
The origin of the path $s$ is chosen at $Z$=4.0 \AA ~above the surface.
The barrier height is 1.55 eV per single H$_2$ molecule, which is comparable
with the earlier calculated value of ${\sim}$1.2 eV per single H$_2$ molecule
given by Hammer and N{\o}rskov\cite{hamm2} calculated using an
earlier version\cite{gga} of the GGA-PBE\cite{pbe} functionals that we used.
Barrier heights have been recognized to be sensitive to electron-correlation
functionals.\cite{hamm6,hamm4,mart}

To investigate the origin of the barrier, we first evaluate the electron
transfer from the two H atoms in the dissociated H$_2$ molecule using the
Bader analysis.\cite{bader}
In order to show the electron population variation, we define the equation
\begin{equation}
  \varDelta x_{2\mathrm{H}} = x_{2\mathrm{H}}[\mathrm{H}_2,\mathrm{Au}(111)] - x_{2\mathrm{H}}[\mathrm{H}_2],
\end{equation}
where the first term is the electron population of the two H atoms in the whole
system and the second term is that in the supercell with the same size as the
first term.
Figure \ref{fig4} shows the electron population variation
$\varDelta x_{2\mathrm{H}}$ along the reaction path $s$ together with the
potential energy variation $\varDelta E$ shown in Fig. 3.
The electron variation of the two H atoms decreases with increasing energy barrier
indicating the electron transfer to the Au substrate.
The path $s$ at the minimum of the population of the dissociate H$_2$ 
coincides with $s$ at the maximum energy $\varDelta E$.
The maximum electron transfer from the H$_2$ molecule to the surface has
been 0.129$e$ per molecule.
Thus the maximum barrier height of the energy corresponds to the most
depleted electron state of the dissociation of the H$_2$ molecule, 
indicating that a correlation exists between the barrier formation
and the electron transfer.

Figure \ref{fig5} shows the variation in $\varDelta E$ as a function
of the difference electron population $\varDelta x_{2\mathrm{H}}$,
in which we used the quantities shown in Fig. \ref{fig4}.
The energy increases linearly with decreasing electron variation in the dissociated
H$_2$ molecule, deviates from the increase at -0.087$e$ shown by an
open square mark, and increases to the barrier maximum shown by an
open triangle.
The slope is -9.37 eV/$e$.
The electron transfer from the molecule to the Au surface is proportional
to the energy increase of the system in the region.

However, a question arises: how can this barrier $\varDelta E$ be explained.
We have just found that the dissociated H$_2$ molecule transfers
the electron to the Au substrate.
Here, we examine how the barrier is related to the energy of isolated,
separated, and charged H$_2$ molecules.
To estimate the energy, we evaluate the binding energy of the
molecules by the real-space density functional method.
By fixing both the charge states of the dissociated H$_2$ molecules given
in Fig. \ref{fig4} and their bond lengths $d$ given in Fig. \ref{fig2},
we calculate the energies of the molecules in real-space.
Figure \ref{fig6} shows the calculated binding energy $E_b$ of the
molecules as a function of the path $s$, together with the energy
$\varDelta E$ of the whole H$_2$/Au(111) system.
The energy $E_b$ increases with the path $s$ and coincides with the
energy $\varDelta E$ curve almost up to the open square mark,
shown in Figs. \ref{fig2} and \ref{fig5}.
This implies that the increase in the energy $\varDelta E$ of the
whole system correlates to the destabilization of the dissociated molecule.
The energy of the interaction between the dissociated H$_2$ molecule
and the Au surface is negligible in this region. 

Figure \ref{fig7} shows the difference electron population
$\varDelta x_{\mathrm{Au}{\sharp}}$ of each Au$\sharp$ atom on the first layer of the slab.
Here we define the difference electron population $\varDelta x_{\mathrm{Au}{\sharp}}$
using the equation
\begin{equation}
  \varDelta x_{\mathrm{Au}{\sharp}} = x_{\mathrm{Au}{\sharp}}[\mathrm{H}_2,\mathrm{Au}(111)] - x_{\mathrm{Au}{\sharp}}[\mathrm{Au}(111)],
\end{equation}
where $\sharp$ is the positions of each Au atoms in the slab.
We also show in this figure the variation of the electron population
$\varDelta x_{2\mathrm{H}}$ in Fig. \ref{fig4}
and the energy variation $\varDelta E$ in Fig. \ref{fig3} for comparison. 
On the one hand, the electron of the top Au0 atom in the slab decreases
with increasing path $s$.
The electron variation increases after the minimum -0.053$e$.
On the other hand, the electron variations of the other coordinated
Au1, Au2, and Au3 atoms on the first layer increase with the path $s$
and decrease after their maxima.
The variations of the Au2 and Au3 atoms are the same, since
they are symmetrical with respect to the two H atoms, as is shown in
Fig. \ref{fig1}(a).

The path $s$ at the maxima of the populations for the Au1, Au2, and Au3 atoms
coincides with that at the minimum population for the Au0 atom.
At this turning point, the deficit population of the top Au0
atom is 0.053$e$.
The deficit electron population of the dissociated H$_2$ molecule at
the point $s$ amounts to 0.087$e$. 
The coordinate position that corresponds to the point is
($d$, $Z$)=(1.0\AA ,1.7\AA), which is shown in Figs. \ref{fig2},
\ref{fig5}, and \ref{fig6} as the open square marks.
The point is far below the energy barrier maximum.
The decrease in the electron variation in the top site Au0 atom is caused by the
Coulomb repulsion between the electrons around the dissociated H$_2$ molecule
and those around the top Au0 atom.
The coordinated Au1, Au2, and Au3 atoms accept the repelled electrons from
the top Au0 atom.
Thus, the electron transfers from the molecule to the top Au0 atom and from
the top Au0 atom to other Au atoms in the first layer.
The path $s$ at the maximum variation for Au0
coincides with the minimum variation for the 2H corresponding
to the energy barrier maximum of $\varDelta E$ in Fig. \ref{fig4}.
The difference electron population of the second layer Au atoms
are negligible compared with those of the first layer.

\section{Discussion}
We have used both the real-space DFT method and the periodic DFT method to 
identify the origin of the barrier height. 
The real-space method uses a lower number of basis functions comparing
with the number of basis functions in the periodic methods.
The real-space methods have a lower accuracy in the electronic structure
calculation than the periodic methods.
The magnitude of the barrier, however, has been large enough and amount to 1.55 eV.
So the incompleteness of the wavefunctions in the real-space method would be 
negligible in the present study.

We have evaluated the potential energy barrier during the dissociative
adsorption of the H$_2$ molecule to the Au(111) surface.
The electron in the dissociated H$_2$ molecule has transferred from the
H$_2$ molecule to the surface during the energy barrier formation. 
The energy curve along the reaction path has coincided with that of the
charged, separated, and isolated H$_2$ molecule that has been
calculated by the real-space method.
They have coincided up to the turning point as shown in Fig. \ref{fig6}.
So the energy barrier is formed by the destabilization of the dissociated
molecule in the region.
The reaction path $s$ at the minimum of the difference electron population
in the dissociated H$_2$ molecule has coincided with that $s$ at the maximum
potential energy $\varDelta E$.
The energy increase has been proportional to the electron transfer from the
dissociated H$_2$ molecule to the Au surface up to the turning point.
The barrier formation is due to the destabilization of the dissociated molecule.

The present results have shown that the electron in the molecule transfers
to the Au(111) surface, although the Au surface has the filled $d$-band.
Even the Au surface with the filled $d$-band has received the electron
from the dissociated H$_2$ molecule.

There may exist two other contribution to the potential energy variation 
for the dissociation of H$_2$ molecule on the Au substrate;
they are the energy contribution from the interaction between the molecule
and the substrate and the one from the electron increase in the substrate.
At present it is unclear whether they are cancelled each other or they are
negligible in the magnitudes.

Here we discuss the origin of the transfer that has caused the barrier
formation.
There were a few earlier studies on the electron transfer during the
dissociative adsorption of the molecules on the metal surfaces,
although these studies gave no relation between the transfer and the energy
barrier formation.
Bird $et ~al.$ showed the increase in electron variation in the dissociated
H$_2$ molecule on the Mg(0001) surface,\cite{bird}
showing that the electron transferred from the Mg surface to the H$_2$
molecule.
Huda $et ~al.$ indicated that the electron transferred from the Pu(111)
surface to the dissociated H$_2$ molecule.\cite{huda}
In the present case, in contrast, the electron has transferred from
the H$_2$ molecule to the Au surface. 
This is in the reverse direction of the transfers in the earlier two cases.
However these directions can be explained consistently by the
electronegativity differences between the H atoms and the metal substrates.
Since the electronegativity is larger for H than for Mg, the electron
transferred from the Mg surface to the H atoms.  
The electronegativity is larger for H than for Pu, so the electron
transferred from the Pu surface to the H atoms.
The electronegativity is larger for the Au than for the H, so the electron
transferred from the H atoms to the Au surface.
Thus the electronegativity difference between the H atom and the metal
substrates determines the direction of the transfer.

Figure \ref{fig2} has shown the maximum barrier in the $\varDelta E$ to appear at
($d$,$Z$)=(1.5\AA ,1.5\AA ).
The distance $d$ is fairly larger than the equilibrium distance 0.74
\AA ~of the H$_2$ molecule, which we call dissociative adsorption.
This indicates that when the H$_2$ molecule approaches the Au surface,
the difference in electronegativity induces electron transfer.
Then the H$_2$ molecule has dissociated owing to an electron deficit that
raises the energy of the positively charged H$_2$ molecule.
When the H$_2$ molecule conserves the equilibrium distance, the energy
variation is higher than that of the reaction path $s$, as shown
in Fig. \ref{fig2}.

For H$_2$ molecule adsorption on Pt surfaces, there has been no
barrier or low barriers in the potential energy surfaces.\cite{olse,hamm5}
The H atoms have been adsorbed with almost the same interhydrogen
distance as that in the equilibrium H$_2$ molecule, i.e., without dissociating
the molecule,\cite{olse,hamm5} which we call nondissociative adsorption.
This is because these elements have the same electronegativity. 
This is another evidence of the origin of the barrier formation found in the
present investigation.

The energy increase of the whole system has coincided with that of the
separated, charged, and isolated H$_2$ molecule up to the turning point,
as shown in Fig. \ref{fig4}.
The extent of the coincidences will depend on the adsorption sites on
the Au substrates, since the distance between the H$_2$ molecule and
the substrate depends on adsorption sites.

We have investigated the electron transfer from the H$_2$ molecule,
whose axis is kept parallel to the surface, to the Au substrate. 
We predict the extent of the transfer from a vertically oriented H$_2$
molecule to the substrate.
Since the electron has transferred from the H$_2$ molecule to the Au
substrate, we expect that the lower H atom transfers more electrons
than the upper H atom to the substrate.
An unsymmetrical electron transfer will cause a higher energy barrier
than the symmetrical case when we have calculated.

The electron transfer found in the present investigation may be related to
Pauli repulsion. 
The investigation of the relation is beyond the scope of the present study.

\section{Conclusions}
We have investigated the electron transfer from the dissociatively adsorbed
H$_2$ molecule to the Au(111) surface by evaluating the electron that
belongs to atoms in crystals and molecules.
We have found that electrons transfer from the molecule to the surface.
The difference of the electronegativities determines the direction of the
electron transfer not only in the H$_2$/Mg(0001) and H$_2$/Pu(111) systems,
but also in the present H$_2$/Au(111) system.
The calculated dissociation energy curve along the reaction path has
coincided with that of the isolated, separated, and positively charged H$_2$
molecule that has been calculated by the real-space density functional method.
The energy increase has been proportional to the charge deficit of the
H$_2$ molecule up to the turning point.
The barrier formation in the present system is due to the destabilization
of the dissociated molecule induced by the transfer.

\section*{Acknowledgments}

The authors thank Professor Masahiro Yamamoto of Kyoto University for
discussions.
The numerical calculations were partly carried out using SCore cluster system
in Meiji University and Altix3700 BX2 at YITP in Kyoto University.

Fig. 1. ~Schematic drawing of the surface structure and the
adsorption sites of hydrogen molecule H$_2$ on the Au(111) surface. 
(a) The periodic supercell consists of four Au layers each containing four atoms
with a $\sqrt{3}\times2$ structure sandwiched by two vacuum layers.
The dashed line corresponds to the unit cell.
The closed small circles indicate the H atoms, and the open circles the Au atoms.
The numbers 0 to 3 correspond to Au0, Au1, Au2, and Au3 atoms, respectively.
(b) Side view intersected at the dotted plane in the left figure.
The distance $d$ is the separation between the two H atoms and the distance $Z$
is the height of the molecule.

Fig. 2. ~Contour plot for the potential energy surface for the dissociation of
the hydrogen molecule H$_2$ on the Au(111) surface as a function of the H-H bond
length $d$ and the height $Z$ of the molecule above the surface.
The contour spacing is 0.1 eV.
The top of the potential energy barrier appears at ($d$,$Z$)=(1.5\AA ,1.5\AA ),
shown as an open triangle mark. 
The open square mark shown at ($d$,$Z$)=(1.0\AA ,1.7\AA ) is referred in the text.
The equilibrium adsorption site of H atoms is located at ($d$,$Z$)=(2.8\AA ,1.1\AA ).

Fig. 3. ~Potential energy variation $\varDelta E$ along the
reaction path $s$ of the hydrogen molecule H$_2$ on the Au(111).
The origin of the $s$ is chosen to be $Z$=4.0 \AA ~above the center of the first
layer of the Au surface. The lines are for visual guidance.

Fig. 4. ~Difference electron population $\varDelta x_{2\mathrm{H}}$
along the reaction path $s$ in H$_2$/Au(111) system together with the potential
energy variation $\varDelta E$ shown in Fig. \ref{fig3}.
The maximum population $\varDelta x_{2\mathrm{H}}$ is -0.129$e$ and the maximum barrier
is 1.55 eV.
The lines are for visual guidance.

Fig. 5. ~Potential energy variation $\varDelta E$ as a function of the
difference electron population $\varDelta x_{2\mathrm{H}}$.
The open square mark at the beginning of the deviation corresponds
to the square mark shown in Fig. \ref{fig2}. 
The open triangle mark corresponds to the maximum barrier height of
the potential energy.

Fig. 6. ~The variation in the binding energy $E_b$ of the isolated,
separated, and charged H$_2$ molecule as a function of the reaction path
$s$, together with the potential energy variation $\varDelta E$ shown in
Fig. \ref{fig3} for comparison.
The binding energies (solid line) are calculated with the same bonding
distances and charged states as those of the H$_2$ molecule in the whole
H$_2$/Au(111) system shown in Fig. \ref{fig4}. 
Both the energies are plotted with the same scales.
The open square mark corresponds to the same mark shown in Fig. \ref{fig5},
and the open triangle to the barrier top.
The lines are for visual guidance.

\begin{figure}
\begin{center}
\begin{overpic}{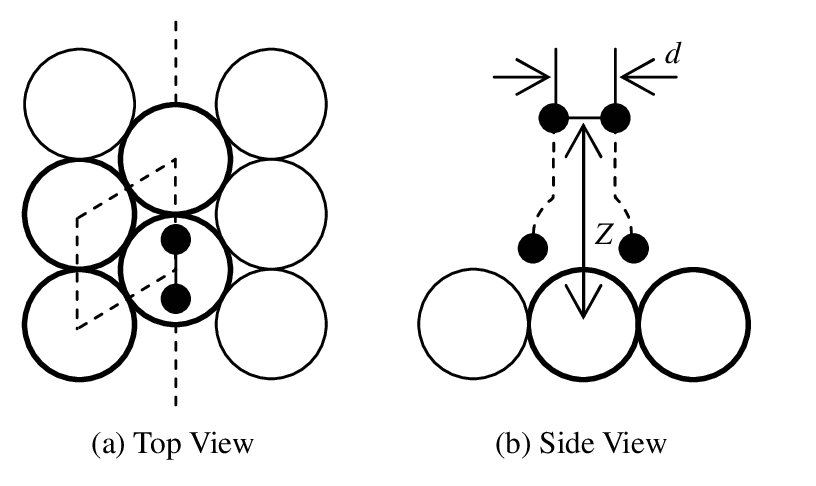}
\put(55,57){0}
\put(55,89){1}
\put(15,39){2}
\put(15,72){3}
\end{overpic}
\end{center}
\caption{Schematic drawing of the surface structure and the
adsorption sites of hydrogen molecule H$_2$ on the Au(111) surface. 
(a) The periodic supercell consists of four Au layers each containing four atoms
with a $\sqrt{3}\times2$ structure sandwiched by two vacuum layers.
The dashed line corresponds to the unit cell.
The closed small circles indicate the H atoms, and the open circles the Au atoms.
The numbers 0 to 3 correspond to Au0, Au1, Au2, and Au3 atoms, respectively.
(b) Side view intersected at the dotted plane in the left figure.
The distance $d$ is the separation between the two H atoms and the distance $Z$
is the height of the molecule.}
\label{fig1}
\end{figure}

\begin{figure}
\begin{center}
\begin{overpic}{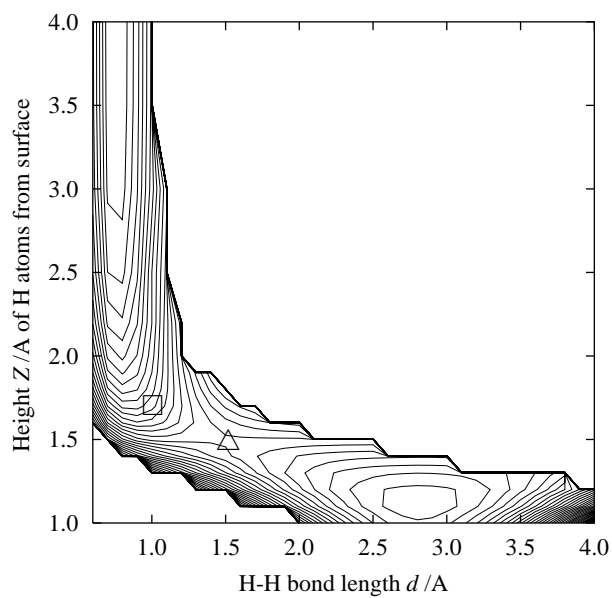}
\put(82.5,55){${\triangle}$}
\put(54.5,68){${\square}$}
\end{overpic}
\end{center}
\caption{Contour plot for the potential energy surface for the dissociation of
the hydrogen molecule H$_2$ on the Au(111) surface as a function of the H-H bond
length $d$ and the height $Z$ of the molecule above the surface.
The contour spacing is 0.1 eV.
The top of the potential energy barrier appears at ($d$,$Z$)=(1.5\AA ,1.5\AA ),
shown as an open triangle mark. 
The open square mark shown at ($d$,$Z$)=(1.0\AA ,1.7\AA ) is referred in the text.
The equilibrium adsorption site of H atoms is located at ($d$,$Z$)=(2.8\AA ,1.1\AA ).}
\label{fig2}
\end{figure}

\begin{figure}
\begin{center}
\includegraphics{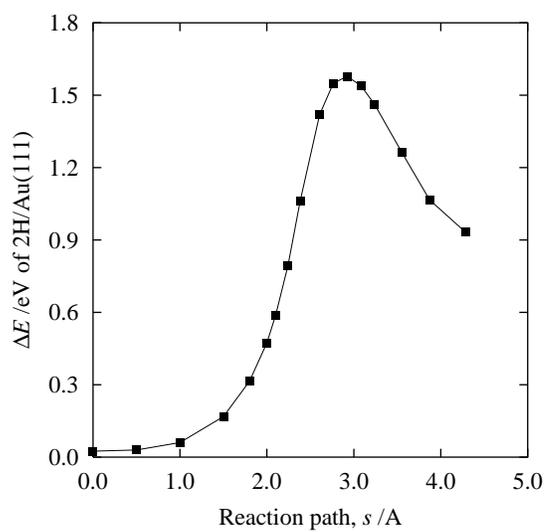}
\end{center}
\caption{Potential energy variation $\varDelta E$ along the
reaction path $s$ of the hydrogen molecule H$_2$ on the Au(111).
The origin of the $s$ is chosen to be $Z$=4.0 \AA ~above the center of the first
layer of the Au surface. The lines are for visual guidance.}
\label{fig3}
\end{figure}

\begin{figure}
\begin{center}
\includegraphics{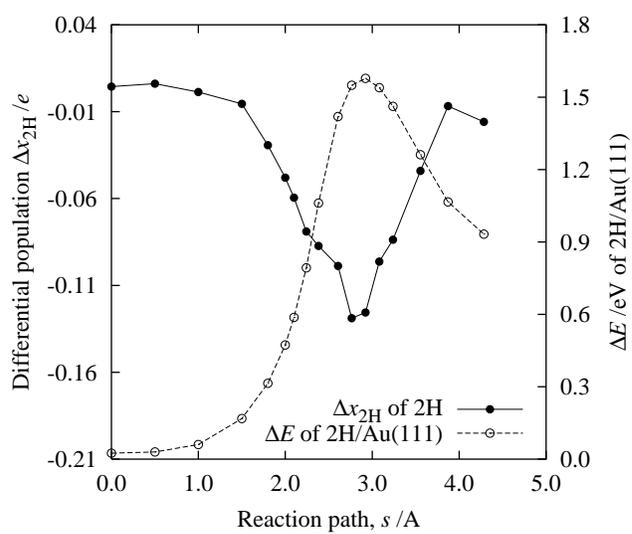}
\end{center}
\caption{Difference electron population $\varDelta x_{2\mathrm{H}}$
along the reaction path $s$ in H$_2$/Au(111) system together with the potential
energy variation $\varDelta E$ shown in Fig. \ref{fig3}.
The maximum population $\varDelta x_{2\mathrm{H}}$ is -0.129$e$ and the maximum barrier
is 1.55 eV.
The lines are for visual guidance.}
\label{fig4}
\end{figure}

\begin{figure}
\begin{center}
\begin{overpic}{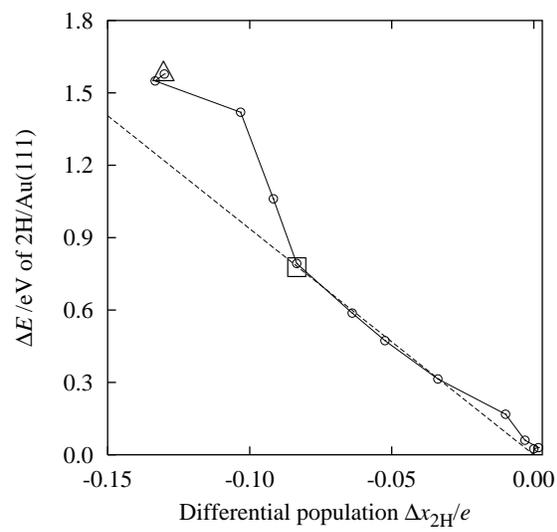}
\put(57,169.5){${\triangle}$}
\put(108,95.5){${\square}$}
\end{overpic}
\end{center}
\caption{Potential energy variation $\varDelta E$ as a function of the
difference electron population $\varDelta x_{2\mathrm{H}}$.
The open square mark at the beginning of the deviation corresponds
to the square mark shown in Fig. \ref{fig2}. 
The open triangle mark corresponds to the maximum barrier height of
the potential energy.}
\label{fig5}
\end{figure}

\begin{figure}
\begin{center}
\begin{overpic}{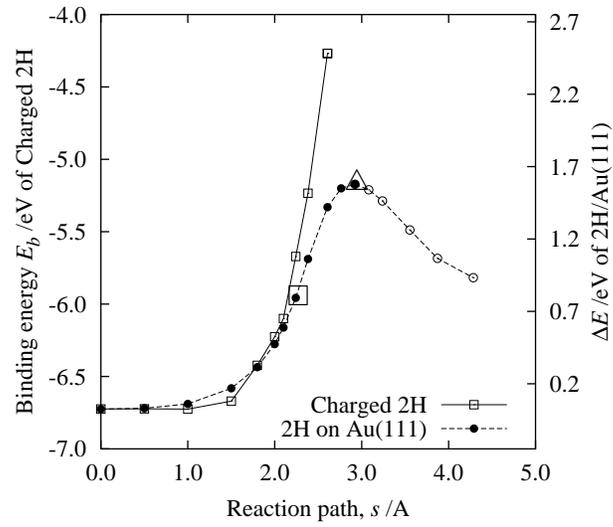}
\put(129,125){${\triangle}$}
\put(107.5,81.5){${\square}$}
\end{overpic}
\end{center}
\caption{The variation in the binding energy $E_b$ of the isolated,
separated, and charged H$_2$ molecule as a function of the reaction path
$s$, together with the potential energy variation $\varDelta E$ shown in
Fig. \ref{fig3} for comparison.
The binding energies (solid line) are calculated with the same bonding
distances and charged states as those of the H$_2$ molecule in the whole
H$_2$/Au(111) system shown in Fig. \ref{fig4}. 
Both the energies are plotted with the same scales.
The open square mark corresponds to the same mark shown in Fig. \ref{fig5},
and the open triangle to the barrier top.
The lines are for visual guidance.}
\label{fig6}
\end{figure}

\begin{figure}
\begin{center}
\begin{overpic}{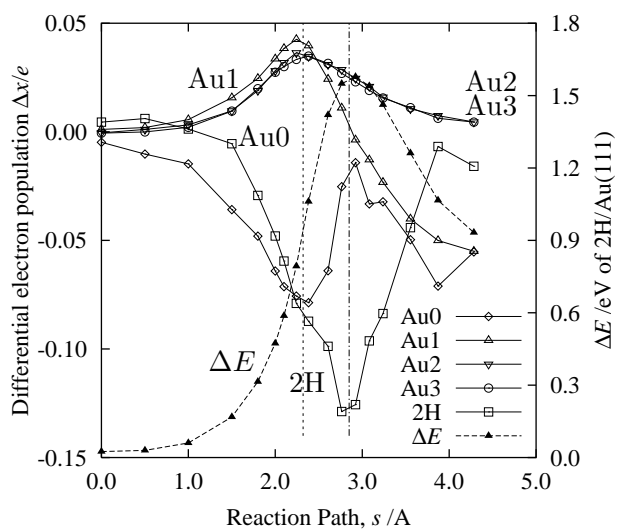}
\put(94,145){Au0}
\put(75,167){Au1}
\put(180,166){Au2}
\put(180,157){Au3}
\put(113.5,52){2H}
\put(83,60){${{\Delta}E}$}
\end{overpic}
\end{center}
\caption{Difference electron population $\varDelta x_{\mathrm{Au}x}$ (solid lines) of each
Au atom on the first layer of the slab structure along the reaction path $s$.
The positions of the Au0 to Au3 atoms have shown in Fig. \ref{fig1}. 
The energy variation $\varDelta E$ (dashed line) and differential electron
population of the two H atoms are also shown for comparison.
The vertical dashed line (left) corresponds to the reaction path $s$ that
corresponds to the open square marks shown in Figs. \ref{fig2}, \ref{fig5},
and \ref{fig6}.
The chained line (right) corresponds to the maximum barrier height of the
potential barrier in the earlier figures.
The lines are for visual guidance.}
\label{fig7}
\label{lastpage}
\end{figure}

Fig. 7. ~Difference electron population $\varDelta x_{\mathrm{Au}x}$ (solid lines) of each
Au atom on the first layer of the slab structure along the reaction path $s$.
The positions of the Au0 to Au3 atoms have shown in Fig. \ref{fig1}. 
The energy variation $\varDelta E$ (dashed line) and difference electron
population of the two H atoms are also shown for comparison.
The vertical dashed line (left) corresponds to the reaction path $s$ that
corresponds to the open square marks shown in Figs. \ref{fig2}, \ref{fig5},
and \ref{fig6}.
The chained line (right) corresponds to the maximum barrier height of the
potential barrier in the earlier figures.
The lines are for visual guidance.
\pagebreak

\clearpage

\clearpage

\clearpage

\clearpage

\clearpage

\clearpage

\end{document}